# Small angle x-ray scattering experiments of monodisperse samples close to the solubility limit

Erik W. Martin[a], Jesse B. Hopkins[b], Tanja Mittag[a1]

[a] Department of Structural Biology, St. Jude Children's Research Hospital, Memphis TN

[b] The Biophysics Collaborative Access Team (BioCAT), Department of Biological Sciences, Illinois Institute of Technology, Chicago, IL

[1] Corresponding author: email address: tanja.mittag@stjude.org

**Abstract**

The condensation of biomolecules into biomolecular condensates via liquid-liquid phase separation (LLPS) is a ubiquitous mechanism that drives cellular organization. To enable these functions, biomolecules have evolved to drive LLPS and facilitate partitioning into biomolecular condensates. Determining the molecular features of proteins that encode LLPS will provide critical insights into a plethora of biological processes. Problematically, probing biomolecular dense phases directly is often technologically difficult or impossible. By capitalizing on the symmetry between the conformational behavior of biomolecules in dilute solution and dense phases, it is possible to infer details critical to phase separation by precise measurements of the dilute phase thus circumventing complicated characterization of dense phases. The symmetry between dilute and dense phases is found in the size and shape of the conformational ensemble of a biomolecule – parameters that small-angle x-ray scattering (SAXS) is ideally suited to probe. Recent technological advances have made it possible to accurately characterize samples of intrinsically disordered protein regions at low enough concentration to avoid interference from intermolecular attraction, oligomerization or aggregation, all of which were previously roadblocks to characterizing self-assembling proteins. Herein, we describe the pitfalls inherent to measuring such samples, the details required for circumventing these issues and analysis methods that place the results of SAXS measurements into the theoretical framework of LLPS.

## 1. Introduction

Intrinsically disordered protein regions (IDRs) are a class of proteins which do not adopt stable secondary or tertiary structure (Oldfield & Dunker, 2014; Tompa, 2012; van der Lee et al., 2014). This lack of structure can be thought of as arising from the flattening of the conformational energy landscape which results in a large ensemble of interconverting conformations with similar free energies. Thus, the characterization of the structure of IDRs must account for the ensemble nature of IDRs. Many traditional structural techniques fail to provide information on IDRs; those that do, including NMR spectroscopy and smFluorescence, provide largely ensemble-average or time-average observables.

Important measures of the structural features of IDRs are parameters that report on their global dimensions – quantities that small-angle x-ray scattering (SAXS) is ideally suited to measuring. Assuming a homogeneous, dilute solution, SAXS directly probes the spatial distribution of atoms in the scattering molecule thus providing direct access to the size and shape of the ensemble of conformations. The utility of SAXS in probing the ensemble properties of IDRs is well established and has been reviewed several times (Bernado & Svergun, 2012; Cordeiro et al., 2017; Kachala, Valentini, & Svergun, 2015; Rambo & Tainer, 2011).

Recent developments in cell biology have drawn attention to the physical process of liquid-liquid phase separation (LLPS) as a relevant driver of non-stochiometric molecular assembly and cellular compartmentalization(Banani, Lee, Hyman, & Rosen, 2017; Shin & Brangwynne, 2017). IDRs were suspected early on as providing the driving force for biomolecular phase separation (Kato et al., 2012; Molliex et al., 2015), perhaps due to their qualitative similarity to well characterized aqueous multi-phase systems involving synthetic polymers (i.e., Polyethylene glycol and Dextran). Indeed, disordered regions have been shown to be sufficient for driving the phase transitions of many proteins (A. E. Conicella, G. H. Zerze, J. Mittal, & N. L. Fawzi, 2016; Kato et al., 2012; Molliex et al., 2015; Nott et al., 2015; Patel et al., 2015). Determining what elements within IDRs are providing the adhesive interactions that drive phase transitions has spurred increased interest in the conformational behavior of IDRs (Brady et al., 2017; Burke, Janke, Rhine, & Fawzi, 2015; Martin et al., 2020; V. H. Ryan et al., 2018).

Polymer phase transitions occur if the sum of three attractive potentials, i.e. polymer-polymer, solvent-solvent and polymer-solvent potentials, favors the solvation of polymer molecules by themselves rather than by solvent. This is captured in the Flory-Huggins mean field theory describing phase separating polymers (Flory, 1942; Huggins, 1942). If a polymer is of sufficient length (often termed the infinite chain limit), Flory-Huggins theory predicts that the emergent properties of a solution of many molecules is recapitulated internally in a single polymer; the affinity between monomers in the same polymer is identical to that across polymers (Figure 1A). While IDRs are heteropolymers and of finite length, homopolymer theory often describes their behavior well suggesting that heterogeneous, long-range interactions within IDRs are sufficiently dynamic that they may cancel on the ensemble level (Hofmann et al., 2012; Martin et al., 2016; Schuler, Soranno, Hofmann, & Nettels, 2016). Indeed, as homopolymer theory predicts, current research suggests that the dimensions of IDRs are predictive of their phase behavior (Dignon, Zheng, Best, Kim, & Mittal, 2018; Lin & Chan, 2017; Martin et al., 2020). Therefore, the details of molecular interactions that can be inferred from precise measurements of dilute, dispersed IDRs provides insight into the properties of dense IDR phases. For example, perturbations to the chemical nature of an IDR either via mutation/permutation of the amino acid sequence or via posttranslational modification (phosphorylation, methylation, etc.) impact the single chain protein properties and can be measured by SAXS; these changes are correlated to changes in the shape of the coexistence curve, or binodal, which quantifies the free energy of concentrated protein solutions. As a consequence of this symmetry, quantitative measurements informative of phase behavior can be made in technologically accessible dilute regimes. In the context of SAXS, this means that IDR shape information can be measured without convolution with interparticle interactions inherent in concentrated solutions.

## 2. Resolution of structural information from SAXS

The minimum distance accessible to a SAXS experiment (Figure 1B) is related to the momentum transfer vector, q.

$q = \frac{4\pi \sin(\theta)}{\lambda}$     (1),

where $\lambda$ is the x-ray wavelength (1-2 Å at synchrotron sources) and $2\theta$ is the scattering angle. The resolution increases with increasing angle such that its length scale is $D = \frac{2\pi}{q}$. The typical SAXS experiment provides a resolution range $0.005 \text{ Å}^{-1} < q < 0.5 \text{ Å}^{-1}$, therefore, the highest obtainable resolution is ~10 Å. However, this cutoff is not strictly analogous to the resolution in high resolution structural techniques such as x-ray crystallography or cryo-electron microscopy (Tuukkanen, Kleywegt, & Svergun, 2016). While it is true that SAXS does not provide details at the amino acid level, parameters such as the radius of gyration ($R_G$) can be calculated at precision far exceeding this resolution (Svergun & Feĭgin, 1986). This apparent dichotomy is rooted in the fact that SAXS is a 'low information' technique where only a limited number of parameters can be extracted precisely (Koch, Vachette, & Svergun, 2003).

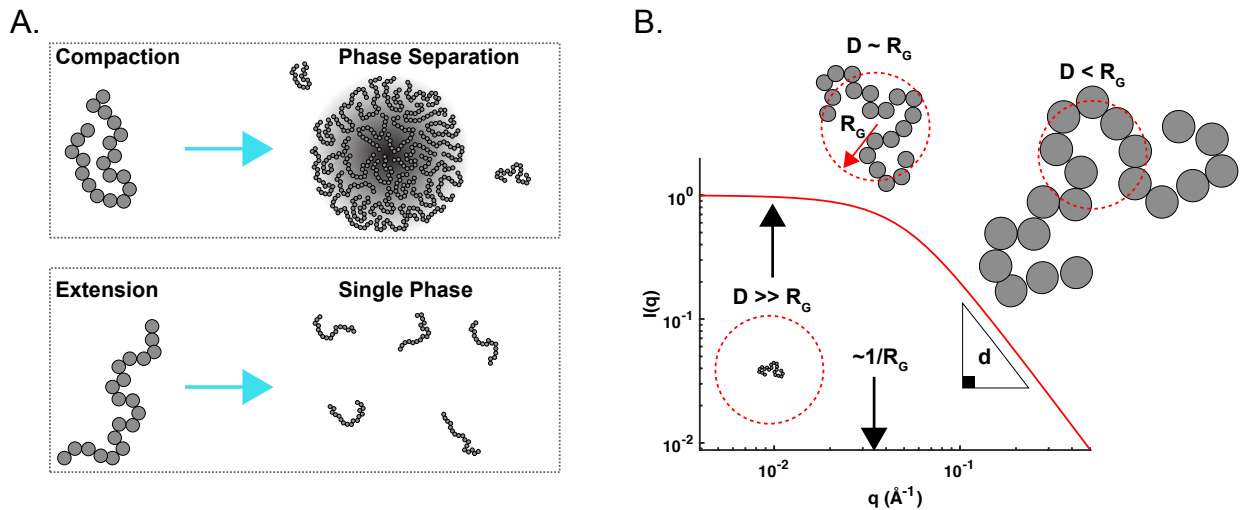

**Figure 1:** Single chain IDR dimensions report on emergent properties. (A) Conditions that lead to dilute chain compaction similarly drive phase separation, while conditions that result in expansion maintain disperse protein. (B) The resolution of a SAXS experiment. The parameter D represents the resolution or distances within the protein that are primarily contributing to the SAXS curve at a defined point. The fractal dimension, d, is defined as the slope in the wider angles.

In addition to the available q range, the resolution of protein SAXS data is limited by the scattering intensity, which is related to the electron scattering length density difference between the protein and the solvent. Protein samples, particularly IDRs, suffer from low scattering contrast that becomes vanishingly small at higher angles. Extending the experiment to wider angles thus does not provide further benefit. As a result, IDR SAXS curves contain no information at the resolution of individual amino acid residues. Completely disordered IDRs contain only two regimes. At wide angles, the scattering decays with a powerlaw, $q^{-d}$, where $d$ is the fractal dimension of the IDR and is typically in a narrow window between ~1.6 and 2. At the smallest angles, the scattering intensity approaches a finite value, $I_0$, that, in the absence of interparticle correlations, is dependent only on concentration and mass. The crossover between the two regimes is defined by a correlation length past which the IDR no longer exhibits fractal-like behavior. This characteristic length is directly related to the $R_G$ of the IDR (Figure 1B)(Hammouda, 1993).

Both $d$ and $R_G$ are ensemble-average quantities in disordered systems and thus report on the entire accessible conformational space of the IDR. Assuming that the sample is monodisperse and sufficiently dilute, $R_G$ can be determined precisely and reflects the mean distance of all protein atoms from the center of mass (Koch et al., 2003). Interpretation of the fractal dimension $d$ can be less straight-forward. If the IDR truly behaves like a homopolymer in the sense that the amino acid positions can be represented by a smooth distribution (i.e., all heterogeneity is averaged out), the fractal dimension can be accurately obtained and should be coupled to $R_G$ via the scaling relation $R_G \sim N^{\frac{1}{d}}$, where $N$ is the number of amino acids in the IDR. The inverse of the fractal dimension is related to the Flory scaling exponent, $\nu = \frac{1}{d}$. These two parameters provide complete information on the characteristics of the conformational ensemble of the IDR if it is statistically random. If the protein properties deviate from statistical randomness as a result of long-range correlations between protein regions, which can arise from strong binary interaction between distant regions, $R_G$ and $\nu$ are decoupled (Banks, Qin, Weiss, Stanley, & Zhou, 2018; Riback et al., 2019). The case of an IDR that is linked to a folded domain is an extreme example of long-range correlations and could have a scattering profile with additional correlation lengths at short distances that reflect the atomic distance distribution in the folded domain. In short, due to the limited resolution of SAXS experiments, the parameters $R_G$ and $\nu$ contain the majority of the information content for samples of completely disordered IDRs. The goal of a SAXS experiment is to accurately extract these parameters to obtain insight into the size and shape of the conformational ensemble and quantify deviations from ideal, random behavior.

## 3. Complications of SAXS measurements of self-assembling proteins

SAXS measurements on soluble biomolecules are straightforward and often done rapidly and efficiently in high-throughput facilities (Classen et al., 2013). However, the very features that allow some IDRs to phase separate limit our ability to make high-quality SAXS measurements. The primary issues fall into two categories. (1) Some phase-separating proteins form soluble oligomers via oligomerization domains which enhance phase separation (Mitrea et al., 2018; Powers et al., 2019; Wang et al., 2018); they may also form off-pathway aggregates (Alexander E. Conicella, Gül H. Zerze, Jeetain Mittal, & Nicolas L. Fawzi, 2016; Molliex et al., 2015; Patel et al., 2015; Schmidt, Barreau, & Rohatgi, 2019). The SAXS curve of a heterogeneous ensemble is the mass average, not the number average, of components, and small populations of large oligomers or aggregates thus have a large contribution to the scattering. This is particularly true at small angles, which are required to precisely determine $R_G$. While SAXS can be used to monitor both oligomerization (Williamson, Craig, Kondrashkina, Bailey-Kellogg, & Friedman, 2008) and aggregation (Herranz-Trillo et al., 2017) both processes severely impair analysis of the ensemble of monomers. (2) The saturation concentration ($c_{sat}$), i.e. the concentration above which the protein forms a dense phase, determines a concentration limit above which SAXS data on monomeric samples is not accessible. However, the effective limit is often lower because the attractive interaction potential between proteins that mediates phase separation can lead to a non-unity structure factor even below $c_{sat}$. Measurements at these low concentrations limit the signal-to-noise of IDR SAXS data with low scattering contrast. These issues are so pervasive that SAXS measurements have historically favored soluble proteins. This point is well illustrated by the fact that the overwhelming majority of IDRs measured by SAXS are more expanded than the polymer theta state (i.e., $\nu > 0.5$), which implies self-avoidance (Bernado & Blackledge, 2009; Cordeiro et al., 2017). These observations have fostered the view that IDRs are generally highly soluble and expanded (Riback et al., 2017). However, phase separating IDRs – under conditions that promote phase separation – have been reported to have $\nu < 0.5$ (Martin et al., 2020) highlighting the ability of adhesive elements in the chain to act inter- as well as intramolecularly.

In order to obtain high-quality data on challenging IDRs characterized by limited solubility, contrast and high aggregation potential, we recommend size-exclusion chromatography (SEC)-coupled measurements at high-flux synchrotron radiation sources. SEC-coupled SAXS eliminates small aggregates and provides flexibility in sample buffer conditions, superior baseline subtraction and rigorous analysis of interparticle interference which we will discuss below.

## 4. IDR sample measurement in SEC-SAXS mode

Ideally, the SEC column will be plumbed to minimize the dead volume between the elution from the column and the x-ray scattering measurement. This minimizes the possibility of sample aggregation or phase separation before its measurement. A critical consideration when developing a SAXS experiment on an associative IDR is to map the conditions under which the IDR is soluble, forms aggregates and phase separates. These conditions will set the practical limits for the experiment. To have the highest obtainable concentration at the x-ray beam, samples need to be loaded onto the SEC column several fold higher in concentration to account for the on-column dilution factor. A practical way to approach this is to load the samples onto the column in a buffer in which the protein is soluble to high concentrations (Riback et al., 2017). For IDRs, this buffer can often contain a denaturant such as guanidinium hydrochloride because there is no risk of perturbing important structural features in the protein in an irreversible manner. Alternatively, buffers that include high (or very low) salt concentrations can be effective in maintaining solubility. The experiment then relies on the buffer-exchanging ability of SEC to place the sample in the correct solution conditions at the x-ray beam.

As a practical consideration, the associative properties of phase-separating proteins can slow their passage through a SEC column. The solubilizing additives can 'push' the protein through the column. This will manifest as one protein peak eluting where expected based on the protein's hydrodynamic radius, followed by a second peak that contains more protein as well as the additives. For this reason, it is useful if the loading has a significantly different conductance than the running buffer to easily distinguish which elution frames contain the sample in the desired buffer.

SEC columns are selected such that resin sufficiently separates the protein from small molecules. SAXS beamlines differ in their preference for silica- or dextran-based SEC resins. Superdex Increase resins from GE Healthcare have been used successfully at the BioCAT beamline at the Advanced Photon Source at Argonne National Lab. An important consideration in column selection is the maximum flowrate. Radiation damage is mitigated by the sample flow and, in a traditional system, the minimum flow rate will be determined by the rate of x-ray damage to the sample – faster flow results in shorter exposure of any given sample volume resulting in less damage. The coflow systems implemented at the Australian National Synchrotron and BioCAT, which will be discussed in detail later, allow for the use of smaller columns and slower flow. A GE Healthcare Superdex Increase 5/150 column with 3 mL volume, which has a maximum flow rate of 0.45 mL / min, is typically sufficient to avoid radiation damage. The minimally allowable flowrate is beamline dependent. While it is impossible to exceed the solubility limit of a particular protein, longer columns allow for larger volumes to be injected; they spread the elution over a larger volume and therefore more frames that can be averaged. The experimental conditions thus represent a tradeoff that has to take into account protein solubility, flux, column flowrate and sample availability.

The SAXS data is collected as individual exposures; their length is a balance between flow rate, the volume in which the sample elutes and the X-ray flux. With a 3 mL SEC column eluting at 0.4 mL / min into a coflow sample chamber, 0.5 second exposures are a good compromise. The resulting data is a series of detector images that can be analyzed in a similar fashion to all SAXS data with freely available software in packages such as BioXTAS RAW (Hopkins, Gillilan, & Skou, 2017) and CHROMIX in ATSAS (Panjkovich & Svergun, 2018).

The obvious benefit of SEC-SAXS is the elimination of oligomers and aggregates (Figure 2). Equally important to the analysis of adhesive proteins are an ideal baseline subtraction (Figure 2) and a built-in continuous concentration series. SEC-SAXS provides the best possible baseline subtraction via the selection of buffer frames that are as close to the protein elution peak as possible. As a result, the composition of the buffer is identical to that in the sample. While this is often a minor consideration when dealing with concentrated samples with high contrast, the quality of the buffer match becomes critically important as the sample concentration decreases. The continuous concentration gradient that is intrinsic in the elution peak from a SEC column allows for critical examination of the data for interparticle interference. Interparticle interference will appear as a dome when the $R_G$ is plotted as a function of elution frame. If associative IDRs oligomerize in a concentration dependent manner on the column, the $R_G$ plot will have a negative slope, because larger species elute earlier from SEC columns. If the association happens with a lag time but is faster than the deadtime between SEC elution and data collection, the $R_G$ plot will also appear as a dome because self-association is strongest at the highest protein concentrations in the middle of the elution peak. Even in the presence of such effects, regions of the SEC elution with the lowest concentration may be selectively averaged to ensure any of these artifacts are eliminated.

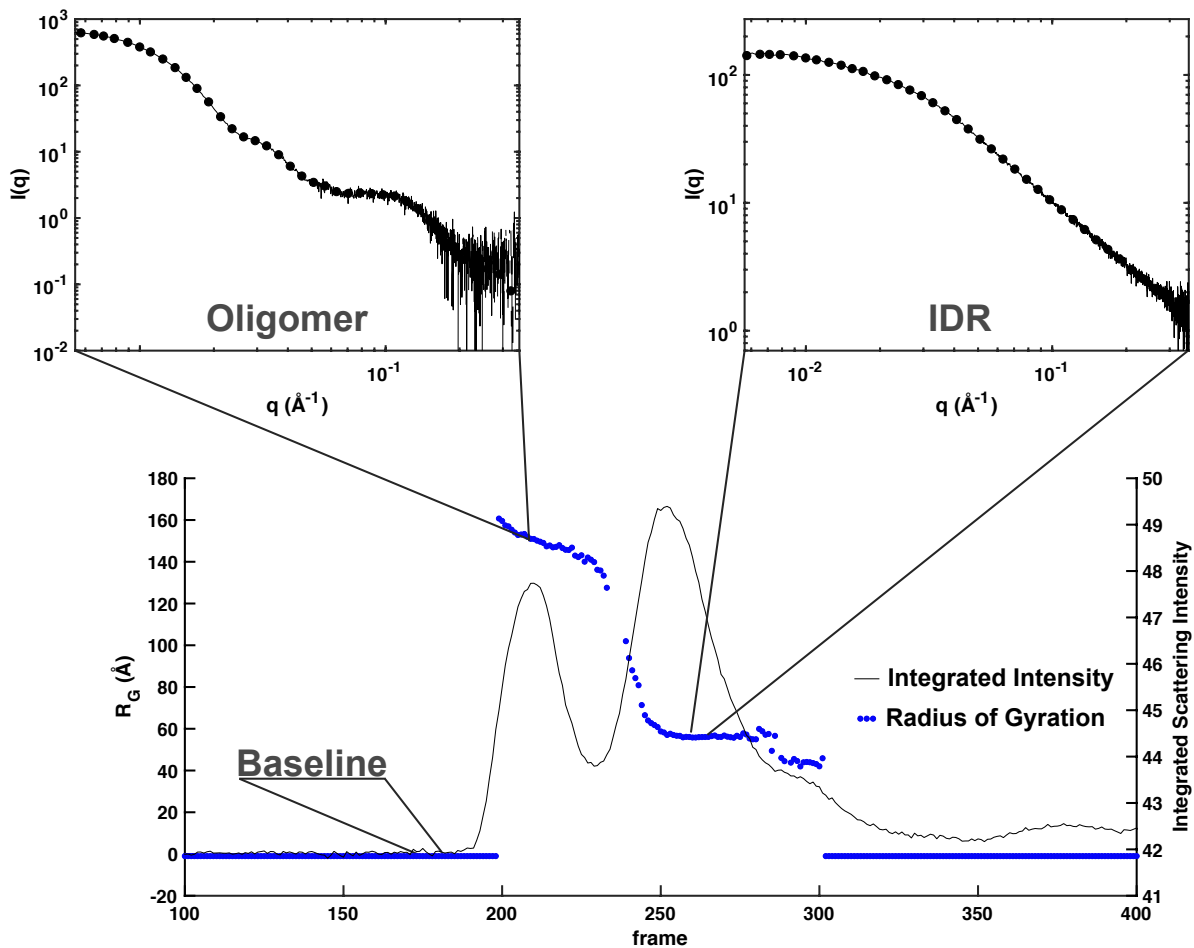

**Figure 2:** SEC-SAXS enables the separation of monomeric IDR from unwanted oligomers and aggregates. The baseline can be chosen from statistically similar frames, where no part of the curve deviates from expected random noise, near the sample elution. The desired sample data is averaged from a region where the $R_G$ is not concentration dependent.

## 5. Primary data analysis

The goal of the SAXS experiment is to extract the ensemble-average radius of the IDR and the shape of the ensemble of conformations. The primary analysis of data on adhesive IDRs proceeds through a series of relatively simple transformations of the scattering profile (Figure 3) which inform on the size and shape.

### 5.1 Guinier analysis
The easiest way to obtain size information from a SAXS profile is a linear fit to the Guinier transform of small angle scattering data. The Guinier equation results from a first order expansion of the Debye scattering equation.

$$I(q) = I_0 e^{\frac{-q^2 R_G^2}{3}} \quad (2)$$

$I_0$, the zero-angle scattering, is determined by protein concentration and molecular weight. In a Guinier plot, data is plotted as the natural logarithm of the intensity as a function of $q^2$ (Figure 3B). Thus, the resulting linear slope is proportional to the square of $R_G$. Due to the fact that the Guinier equation discards all higher-order terms in the expansion, the equation is only valid at very small angles, which for IDRs is taken to be approximately $q < \frac{1}{R_G}$ (Svergun, Feĭgin, & Taylor, 1987).

Guinier analysis provides a good first approximation of the $R_G$ and information on sample quality. The pattern of the residuals in a linear fit of ln(I) versus $q^2$ (Figure 3B) can indicate aggregation or interparticle interference, both of which manifest as an upturn at the smallest angles. While these data points can be excluded, the effects impact the whole scattering profile. When analyzing low-concentration data on IDRs, Guinier analysis can be of limited utility. The noise combined with the limited utilizable q range can result in a high uncertainty of the $R_G$, and these issues are magnified if there is the slightest concern that the sample is not monodisperse.

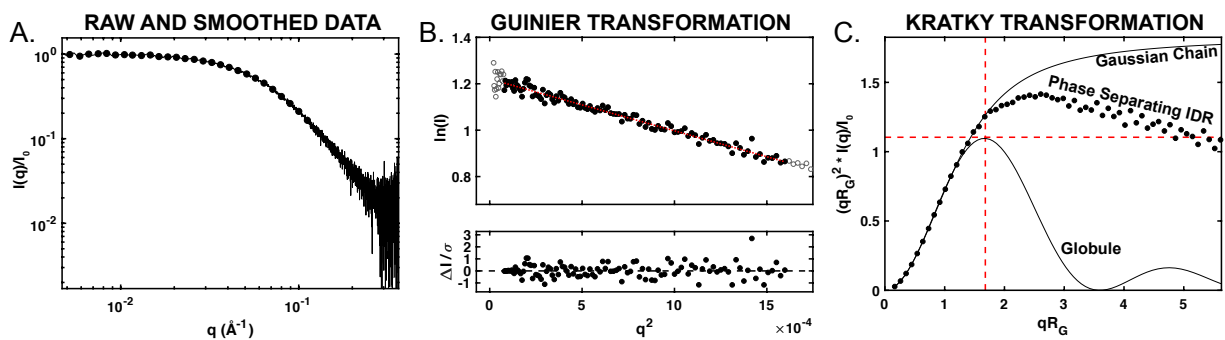

**Figure 3:** Primary SAXS data transformations. (A) Raw SAXS data on IDRs presented in log-log format highlights the low q data, powerlaw regimes and correlation length. Binned data is shown as circles. (B) The Guinier transformation of low q ($qR_G < 1$) data is used to calculate $R_G$. Lower plot shows the residuals. (C) The Kratky transformation enables visual inspection of the fractal dimension as indicated by the slope of the data at high q relative to Gaussian chain and globule references. On a dimensionless Kratky plot, the intersection of red lines indicates the location of the maximum for a globule. Deviation of the maximum from this point indicates flexibility.

### 5.2 Kratky transformation
The Kratky transformation scales the intensity by $q^2$ (Figure 3C). In folded proteins, a Kratky plot can inform on the flexibility of the system. The Kratky plots of well-folded proteins have bell curves with well-defined maxima and converge to zero at higher q (Figure 3C). In contrast, the Kratky plot of unfolded proteins appears hyperbolic and serves as a visual indicator of d. An IDR at the theta point, with *d* = 2, reaches a plateau that continues through the fractal region of the curve, although practically, data quality at higher angles is rarely good enough to visualize the high q regime with high confidence. For self-avoiding IDRs with d < 2, the Kratky plot at high q has a positive slope. For self-interacting IDRs with d > 2, the slope is negative (Figure 3C). A so-called normalized Kratky plot results from normalizing by the zero-angle scattering and $R_G$. By normalizing the intensity by $I_0$, samples of different concentrations or molecular weights can be compared. Multiplying q by $R_G$ normalizes the distance resolution by the protein radius. In the context of a Kratky plot, these normalizations allow for direct comparison of d, and therefore the scaling exponent ν across samples (Durand et al., 2010).

## 5.3 Comments on indirect Fourier transform

In x-ray crystallography (XRC), molecules present in uniform orientation in the crystal lattice and the resulting diffraction pattern represents the position of individual molecules in inverse space. With knowledge of phases, Fourier transform of this data yields the position of atoms in 3D space. Similarly, the SAXS pattern is also determined by the distribution of atoms in space. However, unlike in XRC, the particles in a SAXS sample are typically randomly distributed resulting in spherical averaging of the particle electron density. In the context of IDRs and highly flexible systems, the data represents an additional average over the ensemble of conformations. The quantity derived from a SAXS experiment that is analogous to 3D coordinates in XRC is the atomic pair distribution function, P(r). The SAXS intensity can be directly calculated from the Fourier transform of P(r).

$I(q) = 4\pi \int_0^\infty P(r) \frac{\sin(qr)}{qr} dr$   (3)

In principle, recovering the P(r) distribution from the SAXS pattern should be possible by the inverse Fourier transform. In practice, experimental limitations in measurable scattering angles would require extrapolation to zero angle and wide angles. To circumvent this issue, the indirect Fourier transform (IFT), in which the data is represented by a series of basis functions, was proposed by Otto Glatter (Glatter, 1977).

Similar methods are implemented in most SAXS data analysis software and generally are effective in determining parameters such as the $R_G$ and the shape within the distance resolution of the measurement. This information is used for ab initio shape determination for well-folded systems (Svergun, 1999). However, IFT is problematic for SAXS data analysis of highly flexible, disordered systems. The issue is inherent to finding the solution to the IFT which is an ill-posed inverse problem. The fit therefore relies heavily on regularization to ensure that the solution is smooth and converges to zero. The IFT requires the probability density to converge to zero at zero distance and a defined maximum dimension ($D_{max}$). Finding an appropriate value for proteins with well-defined shape can be accomplished by sampling $D_{max}$ around an estimated solution while monitoring the quality of fit. In the case of IDRs, the $D_{max}$ is poorly defined. Conformations with a $D_{max}$ that approaches the contour length of the IDR may exits but with vanishingly small populations - a problem that is magnified by the weaker contrast for very extended conformations. Because the solution for $R_G$ from IFT depends on the choice of $D_{max}$, the $R_G$ from IFT for IDRs can, at best, have a large error and, at worst, be inaccurate.

The P(r) distribution has some utility in analyzing SAXS data on phase-separating proteins. The shape of the P(r) distribution is likely similar at intermediate distances for all possible solutions and could thus be used to assess changes in shape. If phase separation in a particular protein is initiated by folding, domain movement or rearrangement, the shape of the P(r) distribution will be diagnostic of these changes. However, it is important to keep in mind that the information content of the P(r) distribution is limited by the resolution of the experiment. For small IDRs, high signal beyond q = 0.25-0.3 is rare. Changes in P(r) that suggest shape changes at intermediate distances may thus be caused by artifacts from low signal or suspect baseline subtraction -- a problem equally likely for larger proteins if higher resolution angles are included in the fit but are not free from issues of low signal or baseline subtraction artifacts. IFT is often advertised as a more precise method for obtaining radii compared to Guinier analysis because it uses scattering data from the full q range, but for IDRs, the problems resulting from IFT often outweigh the benefits.

## 6. Synchrotron SAXS beamline hardware

Making high quality measurements of IDRs with low inherent contrast and concentration mandates attention to maximizing signal-to-noise. This requires the use of high-flux synchrotron radiation sources that are specifically designed for biological SAXS measurements.
The Advanced Photon Source (APS) at Argonne National Lab is a high brilliance, third generation synchrotron radiation source ideally suited for measurement of demanding biological samples. The Biophysical Collaborative Access Team (BioCAT) beamline at Sector 18 is dedicated to biological small angle x-ray scattering (SAXS), both of fibers and muscles (fiber diffraction) and solutions and has optimized x-ray optics and detection hardware. The BioCAT beamline is a good model for discussing hardware and will thus be the template for this section. Similar beamlines at synchrotrons around the world will be discussed at the end of the section. The overall layout and primary x-ray optics of the beamline are well described (Fischetti et al., 2004). In brief, BioCAT uses both vertical and horizontal focusing optics to run monochromic SAXS experiments at 12 keV with ~$2*10^{13}$ ph/s in the full 30x140 µm$^2$ focused beam. Collimation for solution SAXS experiments results in an available experimental flux of ~$6*10^{12}$ ph/s. A Pilatus3 X 1M detector (Dectris) is used to measure the resulting scattering.

The signal-to-noise in SAXS measurements can be improved in three ways via the beamline hardware: improving x-ray detection, reducing background scattering, and increasing x-ray flux. X-ray detector technology is already mostly optimized. In the last ~10 years, x-ray pixel array detectors (PADs) (Barna et al., 1995) have become the most common detector type for SAXS. These are single photon counting detectors with zero readout noise, high dynamic range, and detection efficiencies near 100% for x-ray energies in the range of 10-12 keV. Therefore, the noise level is only limited by the Schott noise (Poisson noise) inherent in a counting experiment. The previous and current generation of PADs, most commonly the Pilatus and Eiger detectors made by Dectris (Kraft et al., 2009), provide large detection areas (up to ~300 x 300 mm$^2$), fast readouts (up to 2000 Hz), and minimal (~1 ms) or no deadtime between images. While improvements are possible for certain applications, these detectors are already ideal for SAXS. Beamlines typically have the largest area detector they can afford to maximize the percentage of scattered photons captured. Charge-coupled device (CCD)-based detectors may still be used at some beamlines but are non-ideal for SAXS due to generally slow readouts, non-zero readout noise, and lower detection efficiencies.

SAXS measures x-rays coming from three sources: the sample, the solution, and the SAXS instrument itself. Minimizing the number of x-rays from the instrument can significantly improve the signal-to-noise of a measurement. Instrumental x-ray scattering, also called background or parasitic scattering, is the dominant component of measured x-rays from the instrument, though there may also be a contribution from x-ray fluorescence. Instrumental scattering comes from anything in the beam path, including x-ray optics, x-ray windows, beam monitoring devices, and most significantly air. All SAXS beamlines use at least a partial vacuum environment, which removes air scattering from the measurement. Some SAXS beamlines use in-air sample cells, which increases flexibility for doing experiments with different types of sample holders at the cost of adding scattering from two x-ray windows and a small amount of air. BioCAT and other SAXS beamlines run experiments in a full vacuum environment, including using an in-vacuum sample cell. This removes as much scattering as possible from the x-ray flight path.

Some things are required to be in the beam path, for example x-ray optical elements used to clean and shape the beam before measuring the sample. The scattering from these items is generally divergent from the incident x-ray beam. By placing these items far away from the sample and using one or more sets of slits or pinholes carefully positioned in the beam path, the excess scattering from these items can be almost entirely removed from the measurement. The exact type and layout of these slits or pinholes varies from beamline to beamline, depending on available space. BioCAT uses three sets of single crystal bladed scatterless x-ray slits with both horizontal and vertical blades (Xenocs and JJ X-ray) to collimate the beam and remove excess instrumental scattering. This, combined with minimizing the number of scattering sources by using an in-vacuum sample cell, keeps the intensity of instrumental scattering ~100-fold lower than the solution/sample scattering.

Signal-to-noise levels for a measurement can also be increased by increasing the x-ray flux on the sample. The maximum flux available at a beamline is limited by the x-ray source and beamline design and cannot be easily improved without expensive and intrusive upgrades. However, an additional, practical, limitation on flux is defined by the susceptibility of the sample to radiation damage.(Hopkins & Thorne, 2016; Jeffries, Graewert, Svergun, & Blanchet, 2015) Thus, strategies which minimize radiation damage without compromising flux are crucial to maximizing signal-to-noise. Historically, strategies for reducing radiation damage have involved a combination of continuously flowing the sample through the x-ray beam to spread the incident flux over more sample and the addition of buffer components that act as radical scavengers. While these methods are partially effective, the majority of high-flux beamlines like BioCAT still require attenuation of the incident beam to prevent damage.

Recently, a novel sample cell design, called a coflow cell (Figure 4A,B), was proposed to mitigate issues surrounding radiation damage (Kirby et al., 2016). The coflow cell provides an exterior sheath of buffer 'coflowing' with a central core of sample solution in a laminar flow regime. This flow geometry both minimizes the likelihood of damage to the sample while simultaneously preventing spurious background scattering that can occur from damaged samples adhering to the cell. Using this design, Kirby et al. were able to use an order of magnitude more x-ray flux without damage to the sample, and were able to take full advantage of the high flux beamline (Kirby et al., 2016). Versions of the coflow cell are used at the SAXS/WAXS beamline at the Australian Synchrotron and BioCAT and allow the beamlines to use all available flux without causing radiation damage, which increases signal-to-noise. The increase in flux coupled to decrease in noise allows for measurement at concentrations as low as 0.4 mg/mL (prior to injection onto an SEC column) for a 12 kDa IDR (Figure 3C). This advance has been critical to acquiring interpretable SAXS data on IDRs with low $c_{sat}$.

In the next ~3 years, the APS will be undergoing an upgrade (APS-U), which will increase flux at the BioCAT beamline. BioCAT will also be adding a new multilayer monochromator, which will further increase available flux. While not all of this flux may be useable for SAXS experiments, any portion that is will further improve signal-to-noise.

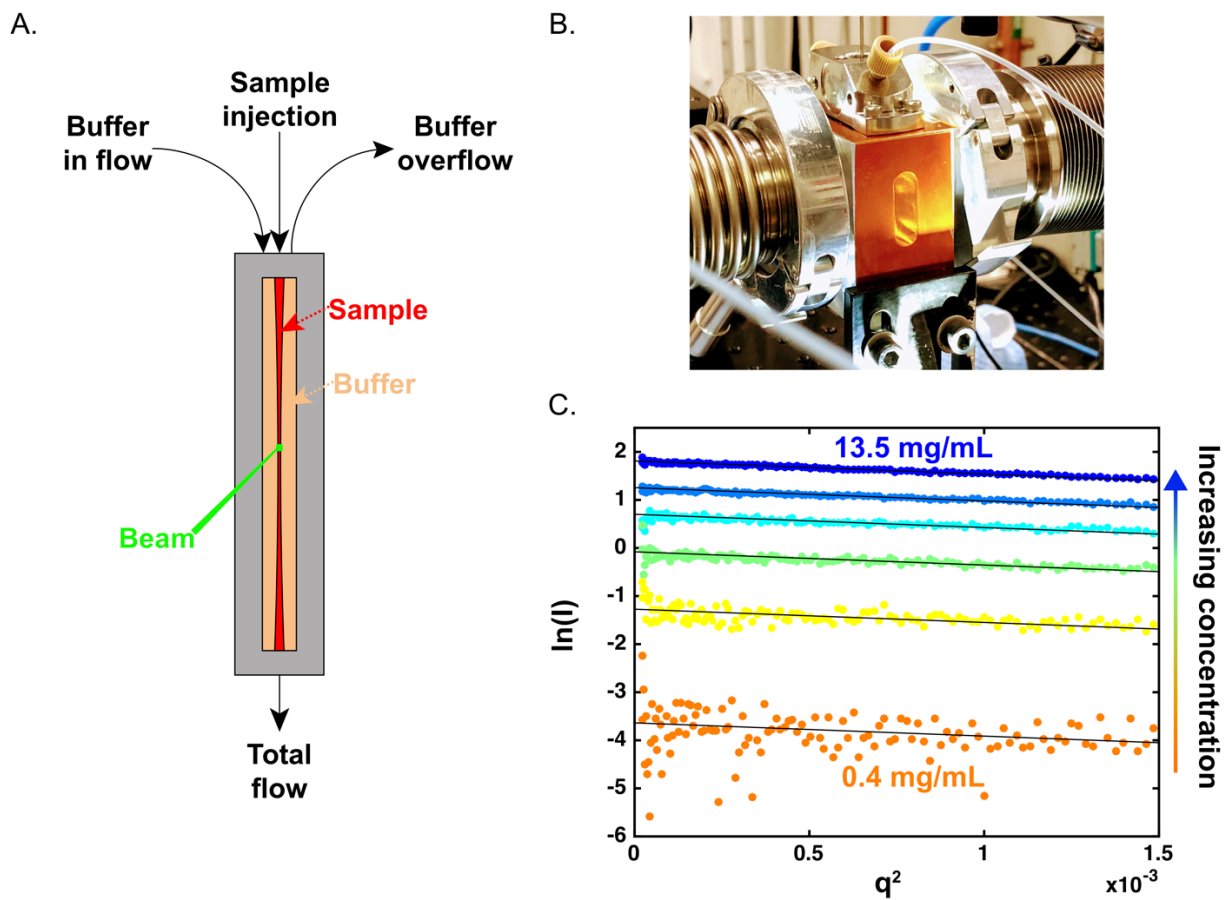

**Figure 4:** The coflow sample chamber enables maximizing signal-to-noise while minimizing radiation damage. (A) A schematic view of the coflow sample chamber. The top of the chamber contains buffer inlet and an outlet to remove excess buffer. The sample is injected into the middle of the capillary where a constant flow is maintained by the total flow outlet pump. The result is the sample 'coflowing' with the buffer in the laminar regime and not mixing. (B) The coflow sample chamber installed at the BioCAT beamline (18ID-D) at the Advanced Photon Source. (C) Low q data shown in Guinier format indicating that identical $R_G$s can be calculated from 0.05 mL samples injected onto a 3 mL SEC column at concentrations ranging from 13.5 to 0.4 mg/mL (12.5 kDa protein).

There are a number of x-ray beamlines around the world with dedicated biological solution SAXS setups similar to that at BioCAT. While details often differ, depending on the specifics of the x-ray source and layout, all beamlines with a dedicated biological SAXS program provide broadly similar resources. These include in-line SEC-SAXS, and increasingly SEC-MALS-SAXS, batch mode SAXS measurements, and often automated sample changing and sample cell cleaning. All beamlines also provide PAD detectors, usually Dectris Pilatus or Eiger detectors with detecting areas equal to or greater than the Pilatus 1M.

Despite the similarities, different beamlines have different specialties. For example, the SIBYLS beamline at the Advanced Light Source (Classen et al., 2013) is highly specialized for high-throughput batch-mode SAXS measurements, using a commercial pipetting robot with custom built sample cells to measure a full 96 well plate every few hours. However, a tradeoff to the focus on high throughput is that the experiments must be done in an air gap, which reduces the signal-to-noise of the measurement. Some BioSAXS beamlines are limited by the source, such as BM29 at the ESRF(Pernot et al., 2013) which is on a bending magnet, limiting the total flux at the beamline to around $10^{12}$ ph/s. Additionally, the adoption of the coflow technology has been somewhat slow, and only two beamlines worldwide currently use the coflow cell (though others are either in the process of or are planning to implement it), which limits the number of high flux beamlines that can fully utilize all available photons. Taken together, there are only a few beamlines particularly suitable for experiments attempting to measure extremely low signal-to-noise systems.

In addition to BioCAT, the other optimal beamline for these measurements is the SAXS/WAXS beamline at the Australian Synchrotron where the coflow system was designed (T. M. Ryan et al., 2018). It has a full beam flux similar to that at BioCAT (~$5*10^{12}$ ph/s), uses an in-vacuum sample cell, and numerous other improvements to minimize excess scattering from beamline components. The P12 beamline at Petra III (Blanchet et al., 2015) and the SWING beamline at Synchrotron SOLEIL (David & Perez, 2009) both provide high flux (up to ~$10^{13}$ and $5*10^{12}$ ph/s respectively), and a fully optimized system with an in-vacuum sample cell and minimal parasitic scattering. However, neither beamline is using a coflow cell, which limits the useable flux and thus the signal to noise achievable. Most other SAXS beamlines around the world lack either the flux or the coflow cell, while several also do not use an in-vacuum sample cell.

## 7. Model dependent analysis

Phase separation of synthetic polymers which are analogous to IDRs has been theoretically modeled for decades. Homopolymer theories either treat the interaction between monomers with a mean field approach (i.e. Flory-Huggins theory) (Flory, 1942; Huggins, 1942) or explicitly consider the networking of monomer units (i.e. Flory-Stockmayer theory) (Flory, 1941; Stockmayer, 1944). In either case, the phase-separating system is considered to be composed of identical monomers distributed in space based on connectivity and entropy. Percolation and networking models can be extended to heterogeneous 'stickers-and-spaces' systems wherein the adhesive elements within the polymer – Flory-Stockmeyer monomers – are the 'stickers' and they are linked by 'spacers' which are similar to non-interacting monomers with positive excluded volume in Flory-Huggins theory. Deducing the adhesive elements in IDRs via careful SAXS experiments can parameterize an IDR system such that it can be treated with an appropriate polymer models with the level of complexity relevant to the system. This section will discuss analytic form factors that can model x-ray scattering from polymer systems. The degree of complexity required to model the single chain system hints at the complexity needed to model phase separation.

When polymer theories are applied to IDRs, it is assumed that IDRs can be treated similarly to ideal polymers or with an additional level of complexity. In this context, it can be useful to interpret SAXS data from IDRs through the lens of polymer models which define the probability density of monomers in the polymer chain based on their covalent connections and non-covalent contacts. Different types of polymer chain models result in different form factors and therefore ensemble size distributions. The form factor P(q) is related to the scattering intensity normalized by $I_0$:
$$I(q) = I_0 P(q) S(q) \quad (4)$$
This form is apparent in equation 2, where the first order expansion of the Debye scattering equation is an estimate of the form factor at small angles. It is important to note that the structure factor S(q) should always equal 1 in dilute solutions. In the presence of interparticle interference, S(q) can take on values below 1 (indicating repulsion) and above 1 (indicating attraction) at small angles. Procedures to model S(q) are well established for a number of colloidal systems but are beyond the scope of this chapter. If these features appear in dilute phase data, it is best to further dilute the sample and only analyze low concentration fractions.

Given the nature of SAXS resolution, all variations of analytical form factors based on the Debye scattering equation (shown in equation 2) should converge on the same $R_G$ at very small angles. Up to which angle – relative to the IDR $R_G$ – the form factor fits the data varies between models. Understanding the assumptions inherent in a model and determining systematic deviations from it, can provide valuable information about the nature of the IDR versus an ideal homopolymer.

### 7.1 Gaussian Chain

The simplest polymer form factor describes a random chain in which the monomers obey Gaussian statistics, i.e. the orientation of each monomer is not correlated with any other monomer. One result of this assumption is that monomers can intersect each other. The Gaussian chain model results in a relationship between the end-to-end distance $R_{ee}$, $R_G$ and number of residues N.

$$\langle R_{ee}^2 \rangle = Nb^2 \quad (5)$$

$$\langle R_G^2 \rangle = \frac{\langle R_{ee}^2 \rangle}{6} \quad (6)$$

This relationship leads to the scaling exponent $\nu = 0.5$. The Kuhn length, b, is the renormalized monomer length that is required so that the IDR can follow Gaussian statistics. Classically, this model would be constructed such that an IDR with N residues is renormalized to N' residues each with a length b such that the model holds. The Kuhn length would then be a function of features of the IDR that result in departure from Gaussian statistics such as steric clashes between sidechains and restrictions to bond rotation. Experimentally, and conveniently, the Kuhn length for IDRs is often ~1 residue (Kohn et al., 2004; Wilkins et al., 1999). The Gaussian chain model allows multiple monomers to occupy the same space. Although this aspect of the model is non-realistic and does not allow the extraction of realistic ensembles of conformations, if the form factor fits the data, the real IDR ensemble will be statistically identical to the non-realistic model. Gaussian chain statistics have the following form factor:

$$P(q) = \frac{2}{q^4 R_G^4} \left[ e^{-q^2 R_G^2} - 1 + q^2 R_G^2 \right]. \quad (7)$$

This form factor was derived by Peter Debye and is a special case of the Debye scattering function where the explicit sum over all atom pairs is replaced by an integral over the probability density (or pair distribution function) of interatom distances in a Gaussian chain. The Gaussian chain form factor has identical limiting behavior as the Debye scattering equation at small angles:

$$P(q) \cong \frac{2}{q^4 R_G^4} \left[ -1 + q^2 R_G^2 + 1 - q^2 R_G^2 + \frac{(q^2 R_G^2)^2}{2} - \frac{(q^2 R_G^2)^3}{6} + \cdots \right] \cong 1 - \frac{q^2 R_G^2}{3} \quad (8)$$

At wide angles, the equations can be simplified to:

$$P(q) \cong \frac{2}{q^2 R_G^2} \sim q^{-2} \quad (9)$$

The fractal dimension of the Gaussian chain, d = 2, is characteristic of the theta state of a polymer where attractive interactions between monomers exactly balance with excluded volume, i.e. repulsive interactions. The theta state can also be thought of as a tipping point above which the polymer is well solvated and below which attractive interactions between monomers are dominant and the polymer collapses. This collapse below the theta state is analogous to phase separation in polymer dense solutions. A consequence of this connection is that an IDR, under conditions where it will phase separate, will likely be near the theta state and therefore a simple Gaussian chain model may be a good fit to experimental data (Figure 5A,B). A single parameter, the $R_G$, describes the Gaussian chain form factor, which is sufficient to describe experimental data sets (equation 7). How far experimental data deviates from the model in the crossover regime and the fractal scaling regime is diagnostic of how similar the conformational ensemble of the IDR is to that of the theta state (Figure 5A,B).

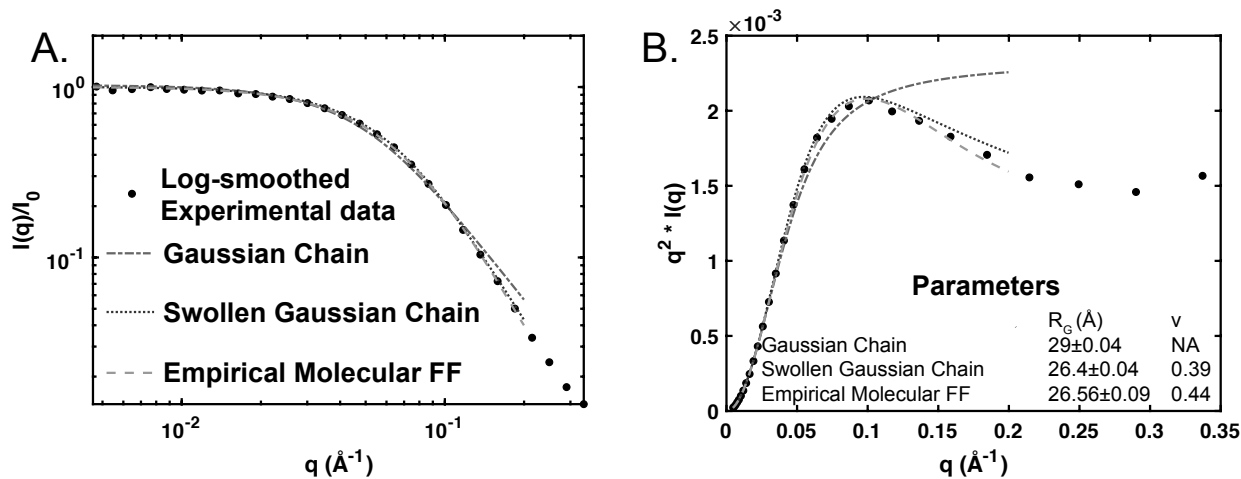

**Figure 5:** Model-dependent fitting of SAXS data on a phase-separating IDR. Fits using the Gaussian Chain, Swollen Gaussian Chain and the Empirical Molecular FF to experimental data on a phase-separating IDR are shown in (A) raw and (B) Kratky format. The Swollen Gaussian Chain and Empirical Molecular FF provide good fits to the data and near equal $R_G$. However, the scaling exponents differ significantly.

### 7.2. Swollen Gaussian Chain

Real polymer chains cannot intersect with themselves, which results in self-avoidance and can be quantified by a parameter called "excluded volume". Excluded volume is added as an exponent to the Gaussian chain statistics from equations 5 and 6:

$$\langle R_{ee}^2 \rangle = N^{2\nu} b^2 \qquad (10)$$

$$\langle R_G^2 \rangle = b^2 \left[ \frac{N^{2\nu}}{(2\nu+1)(2\nu+2)} \right] \qquad (11)$$

where $\nu$ is the Flory scaling exponent which is generally between $\frac{1}{3}$ for collapsed polymers and globules and $\frac{3}{5}$ for self-avoiding polymers. The scaling exponent of $\frac{1}{2}$ represents the theta state and restores the Gaussian statistics in the previous section – a special case where the excluded volume is effectively zero. Unlike for the Gaussian chain, there is no analytic solution to the form factor that includes self-avoidance for any value of excluded volume other than $\frac{1}{2}$. The solution was originally put into integral form by Henry Benoit in 1957 and was more recently put in a more tractable 'semi analytic form' by Boualem (Hammouda, 1993) using a combination of incomplete gamma functions.

$$P(q) = \frac{1}{\nu U^{\frac{1}{2\nu}}} \left[ \gamma\left(\frac{1}{2\nu}, U\right) - \frac{1}{U^{\frac{1}{2\nu}}} \gamma\left(\frac{1}{2\nu}, U\right) \right] (12)$$

$$U = \frac{q^2 R_G^2 (2\nu+1)(2\nu+2)}{6} \qquad (13)$$

$$\gamma(x, U) = \int_0^U e^{-t} t^{x-1} dt \qquad (14)$$

This form factor can be useful for obtaining $R_G$ because it is valid to wider angles than the Guinier approximation or the fit to a Gaussian coil form factor due to the variable parameter $\nu$ (Figure 5A,B). $\nu$ itself is a useful parameter for the characterization of IDR size scaling and can even hint at a driving force for phase separation in some proteins (Martin et al., 2020). However, the absolute accuracy of the parameter $\nu$ extracted by a fit to this form factor has been called into question (Figure 5B) (Riback et al., 2017). Specifically, the swollen Gaussian coil form factor can fail to reproduce parameters of limiting cases such as self-avoiding walks generated by molecular simulations. Given these limitations, it is important to be cautious when interpreting fitted values. Differences in values in response to changes in condition or IDR sequence are likely meaningful, while the absolute values with respect to theoretical limiting cases should be viewed with skepticism.

### 7.3. Empirical derivations

Instead of deriving a form factor based on an ideal, random pair distribution function, an alternate strategy is to empirically derive a form factor based on molecular simulations. This strategy was effectively implemented using Monte Carlo simulations of polystyrene by Pederson and Schurtenberger (Pedersen & Schurtenberger, 1996). In this application, simulations were used to parameterize numerical solutions to the form factor by linear combination of the form factor of a Gaussian chain and an infinitely thin rod. Riback et. al used a similar approach to parameterize a form factor specific to IDRs. They used a series of molecular dynamics simulations in which the excluded volume was effectively titrated by adjusting the attractive potential between beta sidechain carbons in a poly-alanine chain (Riback et al., 2017). The simulations were used to parameterize an empirical function that fits scattering data as a function of $R_G$ and $\nu$ (Figure 5A,B). The empirical molecular form factor (FF) has proven useful in analyzing SAXS data on IDRs (Banks et al., 2018; Martin et al., 2020; Riback et al., 2017; Riback et al., 2019). The IDR-specific parameterization aids in this model's ability to reproduce values of $\nu$ across sequences (Figure 5B). Further, using a model derived from IDRs allows for rapid evaluation of SAXS data relative to an ideal, homopolymeric IDR sequence. Specifically, by evaluating the prefactor, obtained from the form factor fit for a given IDR sequence length, $R_G$ and $\nu$, relative to the prefactor in the poly-alanine model, it is possible to diagnose deviations from random statistics. For example, local stiffness or transient structuring could result in deviations in the Kuhn length which would be manifest as an atypical prefactor.

### 7.4 Simulation-based analysis

Typical samples of synthetic polymers contain a distribution of molecular weights. Assuming that the IDR sample is pure (a feature that should be controlled for and ensured by using SEC-SAXS), knowledge of the chemical identity of the IDR can be used to improve SAXS data interpretation over that with the polymer models from above. Knowledge of the sequence length (as mentioned in the previous section) and amino acid identity can provide details aiding in analysis.

Incorporating information about protein sequence into the analysis usually involves generating an explicit ensemble of conformations that fits the experimental data. A number of different methods exist, but all rely on the concept that the data can be represented by a linear combination of representative conformations. First, an ensemble of conformations is generated, and this can be done by sampling from amino-acid specific degrees of freedom, creating the starting distribution of conformations. The ensemble can also be generated by Molecular Dynamics or Monte Carlo simulations with a physics-based forcefield. Second, the fit of the data to the ensemble is inspected. In the best-case scenario (e.g. in the case of a well-performing force field and sampling method), the starting ensemble gives good agreement with the data. If the fit is deemed suboptimal, the conformations in the ensemble can be reweighted or a sub-ensemble can be selected. This process has been accomplished via Bayesian inference optimization of the complete prior distribution (Antonov, Olsson, Boomsma, & Hamelryck, 2016), genetic algorithms which subsample and reweight the prior (Bernado, Mylonas, Petoukhov, Blackledge, & Svergun, 2007; Pelikan, Hura, & Hammel, 2009), clustering followed by maximum entropy reweighting (Rozycki, Kim, & Hummer, 2011), and reweighting the prior based on minimization of a pseudo-energy through simulated annealing (Krzeminski, Marsh, Neale, Choy, & Forman-Kay, 2013;

Marsh & Forman-Kay, 2012). The quality of the reweighted ensemble strongly depends on the starting ensemble; if the distribution of properties in the starting ensemble is smooth and already recapitulating the experimental data closely (e.g. because the force field performs well), then reweighting has been shown to result in ensembles with physically reasonable properties such as smooth size distributions. If the starting ensemble is in poor agreement with the experimental data, reweighting can result in ensembles with rough distributions of properties that are physically unreasonable. Hence, realistic and extensive sampling of the conformational space of the sequence in question is key.

The low information content of SAXS data results in a danger of overfitting. This is particularly true if the number of conformations in a minimal ensemble are similar to the information content in the SAXS curve, which may erroneously result in suggesting heterogeneity in the ensemble. A solution to preventing overfitting is implemented in the MultiFoXS algorithm which increases the size of the ensemble until it no longer significantly improves the quality of the fit (Schneidman-Duhovny, Hammel, Tainer, & Sali, 2016). This can also be accomplished through Bayesian inference optimization of minimal ensembles (Yang, Parisien, Major, & Roux, 2010). Intriguingly, the majority of SAXS curves of IDRs can be well represented with fewer than five conformations. These results stress the low information content in SAXS data and the fact that useful data analysis should be testing the compatibility with a defined physical model. The molecular details of conformations in minimal ensembles, including the contacts that result in their size and shape and correlations between such interactions, all of which are sought when trying to understand how self-association is encoded in IDRs, cannot be extracted from minimal ensembles. Therefore, the only reliable parameter in minimal models may be the ensemble-average level of compaction – a parameter that is rapidly accessible from analytic models – and not features like intramolecular contacts.

All-atom or coarse-grained simulations using physics-based force fields can in favorable cases be used for the molecular interpretation of SAXS data without fitting to it. The SAXS form factor calculated from an ensemble that represents the conformational space within the simulation can be directly compared with experimental data. This type of approach has been effectively applied using Monte Carlo simulations with the ABSINTH implicit solvent model (Martin et al., 2016; Martin et al., 2020). In case the form factor from simulations and experiment are in agreement, the simulations are likely capturing physically realistic conformational behavior of the IDR. Comparison to additional data, e.g. from NMR experiments, helps to establish whether the simulations capture additional features of the conformational behavior of the IDR such as formation of transient contacts or residual secondary structure.

Molecular-detail insight from simulations is particularly useful when evaluating the conformations of phase-separating IDRs because intra-chain contacts can be assumed to also exist intermolecularly in semi-dilute and dense solutions and drive phase transitions. Thus, SAXS-validated simulations can aid in identifying the adhesive elements in IDRs. The synergy between simulation and experiment can be exploited to devise a mutational strategy in which proposed adhesive elements in an IDR are rapidly probed via simulation and their impact verified by SAXS (Martin et al., 2020).

A well-known difficulty in this approach is finding a simulation paradigm that accurately reflects experimental data – a problem particularly relevant to IDRs for which simulations often result in overly compact ensembles (Best & Mittal, 2010). Continued force field development with particular attention being paid to the interaction between solvent and protein has helped ameliorate these shortcomings in recent years (Best, Zheng, & Mittal, 2014; Robustelli, Piana, & Shaw, 2018). If current forcefields that are optimized for IDRs are insufficient to match experimental data, *ad hoc* modification of the force field or simulation parameters may help. This can often be accomplished by tuning the simulation temperature (Das, Huang, Phillips, Kriwacki, & Pappu, 2016; Francis, Lindorff-Larsen, Best, & Vendruscolo, 2006; Martin et al., 2020) or by subtle modifications to the interactions between solvent and protein (Best et al., 2014; Larsen et al., 2019; Piana, Donchev, Robustelli, & Shaw, 2015). Tuning force field parameters and reweighting ensembles can even be mathematically equivalent. In the context of a mutation-based experimental design, the force field could be tuned to recreate experimental data of an experimentally well-characterized IDR sequence. This same force field is used to simulate sequence variants where proposed adhesive elements are modified. If the force field modifications succeed in capturing physically realistic behavior, the simulation results should be a good fit to experimental data on all variants without any further modification.

An alternative approach is bottom-up coarse-grained simulations in which intraresidue interaction potentials are parameterized to match experimental data (Norgaard, Ferkinghoff-Borg, & Lindorff-Larsen, 2008). These types of simulations can be implemented in generalized coarse-grained engines such as HOOMD-Blue (Anderson, Lorenz, & Travesset, 2008). This analysis is particularly attractive for studying phase-separating systems. Parameterizing a coarse-grained simulation enables testing of models in which the affinities between particular residues or groups of residues, so-called 'stickers', are titrated and the resulting global dimensions are compared to experimental data. The profound advantage of coarse-grained systems is that the same forcefield can be transferred to dense IDR solutions to assay how the interaction potentials translate into emergent properties. An extreme example of this type of simulation are coarse-grained on-lattice Monte Carlo simulations that allow for rapid sampling and convergence in simulations containing many IDR molecules (Choi, Dar, & Pappu, 2019; Holehouse & Pappu, 2020). The PIMMS and LASSI software packages are designed specifically to implement this workflow. PIMMS was used recently to translate the interaction potential between individual amino acids parameterized based on experimental SAXS data into the calculation of complete coexistence curves for a phase-separating IDR (Martin et al., 2020). An alternate approach relies on 'slab geometry' molecular dynamics simulations (Dignon, Zheng, Best, et al., 2018; Dignon, Zheng, Kim, Best, & Mittal, 2018). These methods also allow the direct transfer of a coarse-grained force field from single chain to dense solutions and have the advantage that they are off-lattice and could contain more information about dense phase structure. However, this comes at the cost of higher simulation time.

## 8. Implementation of SAXS measurements of single chain IDR behavior for characterizing phase behavior

Careful analysis of dilute protein data has the potential to reveal vital information about the nature of sequence features that contribute to higher-order assembly and phase separation. Data carefully collected on a series of IDRs with progressively higher aromatic amino acid content displayed a clearly defined drive to collapse (Martin et al., 2020) (Figure 6). The collapse was quantified by the decreasing scaling exponent (Figure 6B) which correctly predicted the phase behavior of these sequences. This example suggests a coherent workflow where IDR sequence or solution conditions can be systematically titrated, and the results directly mapped to protein phase behavior. If the IDR behaves like a homopolymer on the length scales measurable by SAXS, this connection will likely exist. However, as IDRs are heteropolymers of finite length, cases will exist with decoupling and symmetry breaking between single chain and emergent properties. This behavior can be predicted within the 'stickers-and-spacers' formalism recently described by Choi et al. (Choi, Holehouse, & Pappu, 2020). If the 'stickers' are strong enough and the 'spacers' have sufficient excluded volume, it is possible that the 'stickers' on a single chain may not interact with each other. Importantly, even if $R_G$ and $\nu$ do not directly report on $c_{sat}$, the dilute protein ensemble still informs on the characteristics of the dense phase via the excluded volume of spacers and the cumulative strength of interactions between stickers – parameters that can be inferred from the scaling exponent and $c_{sat}$ respectively. Systems where $R_G$ and $\nu$ do not, *a priori*, suggest a low $c_{sat}$, but the stickers are strong enough to drive assembly could result in a lower dense phase concentration or even a percolated, gelled, network in the absence of phase separation (Harmon, Holehouse, Rosen, & Pappu, 2017).

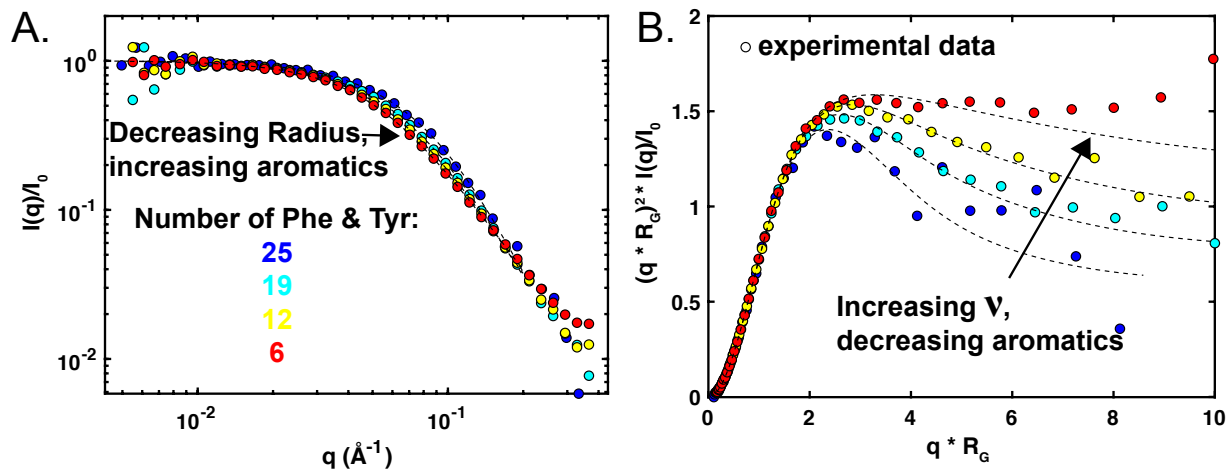

**Figure 6:** Changing sequence features that influence the driving force for phase separation are reflected in the protein dimensions under dilute conditions. (A) Raw data on a series of IDR variants with different numbers of aromatic amino acids shows the predicted decrease in radius with increased propensity to phase separate. (B) The scaling exponent increases with the removal of aromatic amino acids indicating more favorable solvation. Dashed lines are fits to the Empirical Molecular FF.

## 9. Conclusions

SAXS has always been a valuable tool for characterizing IDRs because it provides information on their ensemble-average sizes and size-distributions. However, these measurements have been largely restricted to highly soluble IDRs due to the requirement for aggregate- / oligomer-free samples of high enough concentration to achieve sufficient signal-to-noise. We propose that these limitations resulted in a bias toward samples that behave as if they are in a good solvent ($\nu > 0.5$)(Bernado & Blackledge, 2009; Cordeiro et al., 2017; Riback et al., 2017). While the oligomerization of self-associating IDRs has been monitored by SAXS in individual, interesting examples (Herranz-Trillo et al., 2017), examination of this category of protein in the dilute regime has largely been ignored. Recent advancements in hardware at synchrotron radiation sources along with refined sample preparation has opened the door to SAXS characterization of IDRs that have traditionally been inaccessible.

**Acknowledgement**
The authors would like to thank Thomas Boothby for providing the sample used in Figure 2. T.M. acknowledges funding NIH grant R01GM112846, the St. Jude Children's Research Hospital Research Collaborative on Membrane-less Organelles in Health and Disease and the American Lebanese Syrian Associated Charities. This research used resources of the Advanced Photon Source, a U.S. Department of Energy (DOE) Office of Science User Facility operated for the DOE Office of Science by Argonne National Laboratory under Contract No. DE-AC02-06CH11357. This project was supported by grant 9 P41 GM103622 from the National Institute of General Medical Sciences of the National Institutes of Health. Use of the Pilatus 3 1M detector was provided by grant 1S10OD018090-01 from NIGMS. The content is solely the responsibility of the authors and does not necessarily reflect the official views of the National Institute of General Medical Sciences or the National Institutes of Health.